\documentclass[lettersize,journal]{IEEEtran}
\usepackage{amsmath,amsfonts}
\usepackage{algorithmic}
\usepackage{algorithm}
\usepackage{array}
\usepackage[caption=false,font=normalsize,labelfont=sf,textfont=sf]{subfig}
\usepackage{textcomp}
\usepackage{stfloats}
\usepackage{bbding}
\usepackage{url}
\usepackage{verbatim}
\usepackage{graphicx}
\usepackage{epsfig}%这一句一定要加上  插入矢量图
\usepackage{float}
\usepackage{cite}
\usepackage{color}
\usepackage{xcolor}
\usepackage{multirow}
\usepackage{makecell}
\usepackage{booktabs}
\usepackage{subfloat}
\usepackage{hyperref}
\usepackage[top=0.8in,bottom=0.6in,left=0.4in,textwidth=7.7in]{geometry}
\usepackage{diagbox}
\usepackage{enumitem}
\usepackage{colortbl}  %彩色表格需要加载的宏包
\definecolor{LightCyan}{rgb}{0.88,1,1}
\hypersetup{colorlinks=true, linkcolor=red, anchorcolor=blue, citecolor=green}

\def\BibTeX{{\rm B\kern-.05em{\sc i\kern-.025em b}\kern-.08em
    T\kern-.1667em\lower.7ex\hbox{E}\kern-.125emX}}
\begin{document}

\title{DiffDet4SAR: Diffusion-based Aircraft Target Detection Network for SAR Images\\ }

\author{Jie Zhou, Chao Xiao, Bo Peng, Zhen Liu, Li Liu, Yongxiang Liu, Xiang Li

\thanks{This work was supported in part by the National Key Research and Development Program of China under Grant 2021YFB3100800, and the National Natural Science Foundation of China under Grant 62376283, 62201588, in part by the Key Stone grant (JS2023-03) of NUDT. Corresponding authors: Li Liu, Yongxiang Liu, and Chao Xiao. }
\thanks{The authors are with the School of Electronic Science and Technology, National University of Defense Technology (NUDT), Changsha 410073, China (emails: {zhoujie\_, xiaochao12, liuli\_nudt}@nudt.edu.cn)}}

\markboth{IEEE GEOSCIENCE AND REMOTE SENSING LETTERS, vol. 21, 2024}%
{Shell \MakeLowercase{\textit{et al.}}: DiffDet4SAR}

\IEEEpubidadjcol
\maketitle

\begin{abstract}

Aircraft target detection in synthetic aperture radar (SAR) images is a challenging task due to the discrete scattering points and severe background clutter interference. Currently, methods with convolution-based or transformer-based paradigms cannot adequately address these issues. In this letter, we explore diffusion models for SAR image aircraft target detection for the first time and propose a novel \underline{Diff}usion-based aircraft target \underline{Det}ection network \underline{for} \underline{SAR} images (DiffDet4SAR). Specifically, the proposed DiffDet4SAR yields two main advantages for SAR aircraft target detection: 1) DiffDet4SAR maps the SAR aircraft target detection task to a denoising diffusion process of bounding boxes without heuristic anchor size selection, effectively enabling large variations in aircraft sizes to be accommodated; and 2) the dedicatedly designed Scattering Feature Enhancement (SFE) module further reduces the clutter intensity and enhances the target saliency during inference.  Extensive experimental results on the SAR-AIRcraft-1.0 dataset show that the proposed DiffDet4SAR achieves 88.4\% mAP$_{50}$, outperforming the state-of-the-art methods by 6\%. Code is available at \href{https://github.com/JoyeZLearning/DiffDet4SAR}{https://github.com/JoyeZLearning/DiffDet4SAR}.

\end{abstract}


\section{Introduction}

Synthetic Aperture Radar (SAR) is an active microwave remote sensing imaging technique that possesses all-weather and all-day image acquisition capabilities. It has been widely applied in various fields including terrain mapping, disaster risk monitoring, and ecology \cite{gong2023small}. In these applications, aircraft target detection is a fundamental component in SAR image interpretation, effectively assisting in dynamic
monitoring of critical areas, situational analysis, and emergency rescues. Compared with targets such as vehicles and ships, the complex structure and scattering mechanism of aircraft targets pose significant challenges for accurate detection.

\begin{itemize}[topsep=0pt]

\item[1)] \emph{Diversity of aircraft target sizes:} In SAR images, different types of aircraft targets and images with different resolutions can lead to variations in target sizes. This multi-size target problem creats several issues when using fixed receptive field techniques for feature extraction. As the network deepens, aircraft targets with small sizes or weak backscattering information can be easily missed.

\item[2)] \emph{Discrete scattering points:} The scattering characteristics of aircraft targets are discrete and discontinuous. Incomplete structures and weak correlations between components make it difficult for detectors to achieve accurate predictions.

\item[3)] \emph{Complex scenes and strong background interference:}  In real-world scenarios, the buildings, vehicles, and certain metallic structures around aircrafts exhibit strong scattering characteristics similar to the aircraft targets, making it challenging to detect targets with sufficient accuracy.

% 场景复杂，背景干扰严重：在现实场景中，飞机周围的建筑物、车辆和某些金属设施具有和飞机目标相似的强散射性，这给准确定位和识别目标带来了挑战。

% \item[3)] \emph{Large intra-class differences: }Due to the sensitivity of imaging azimuth, even for the same type of aircraft target, the geometric contour and scattering center presented in SAR images are not entirely consistent, resulting in large intra-class differences and increasing the difficulty of aircraft target fine-grained recognition.
\end{itemize}

Recently, deep learning-based methods with powerful feature learning capabilities have achieved significant progress in SAR image target detection.
These methods are generally based on convolution neural networks and transformers. Using deep separable convolution, Chang \emph{et al.}\cite{chang2023mlsdnet} developed the MLSDNet to obtain the scattering information and aggregate the position and contour features of multi-scale targets. Niu \emph{et al.}\cite{niu2023aircraft} combined the Swin-Transformer with CNN to obtain multi-scale global and local information. Their method overcomes the low signal-to-noise ratio of aircraft target detection in SAR images. Although the abovementioned studies have achieved competitive results, they are somewhat complex and depend on a fixed set of learnable queries. In particular, they use the Feature Pyramid Network (FPN) or prior assumptions to generate region proposals and bounding boxes, requiring prior knowledge and complex network architecture design. Thus, there is an urgent need for new paradigms for SAR image aircraft target detection that achieve satisfactory performance.

Diffusion models have been extremely successful in a wide range of applications, 
including 3D classification\cite{shen2024diffclip}, segmentation\cite{kolbeinsson2024multi}, and object detection\cite{chen2023diffusiondet}. Inspired by \cite{chen2023diffusiondet}, we make the first attempt to apply diffusion models to aircraft target detection in SAR images and propose a \underline{Diff}usion-based aircraft target \underline{Det}ection network \underline{for} \underline{SAR} images (DiffDet4SAR). Specifically, we map the SAR aircraft target detection task to a denoising diffusion process from noisy bounding boxes to ground truth bounding boxes without heuristic anchor size selection. This enables good adaptation to large-size variations in the aircraft within SAR images. Considering the sparse scattering points in aircraft targets and strong background interference, we design a scattering feature enhancement (SFE) module that enhances the integration of scattering features by capturing texture details and semantic information. The use of SEF module effectively highlights the target reception capability. The main contributions of this letter can be summarized as follows:

\IEEEpubidadjcol

\begin{itemize}[topsep=0pt]

\item[(1)] To our best knowledge, this letter presents a first study that introduces diffusion models to SAR target detection, offering a novel, highly simple yet effective framework, dubbed DiffDet4SAR, tailored for SAR target detection.

%  To the best of our knowledge, this is the first attempt to apply diffusion models to the problem of SAR target detection, providing a promising new paradigm for SAR image interpretation.

\item[(2)] The proposed DiffDet4SAR is based on two core designs. First, we cast the SAR target detection problem as a denoising diffusion process from noisy bounding boxes to precise target bounding boxes. Second,  we design a scattering feature enhancement (SFE) module to effectively reduce the scattering intensity of background clutter and highlight targets to mitigate the problems of discrete scattering points of aircraft targets and severe background interference. 

% We design the scattering feature enhancement (SFE) module, which addresses the problems of discrete scattering points of aircraft targets and severe background interference. The SFE module effectively reduces the scattering intensity of background clutter and highlights targets.

\item[(3)] By coupling these two designs, the performance of DiffDet4SAR significantly surpasses the state of the art on the SAR-AIRcraft-1.0 dataset.

% Experimental results on the SAR-AIRcraft-1.0 dataset demonstrate that the proposed DiffDet4SAR outperforms the state-of-the-art methods.
\end{itemize}

\begin{figure*}[t]   
 \centering
  \subfloat{}    
\includegraphics[width=0.62\textwidth]{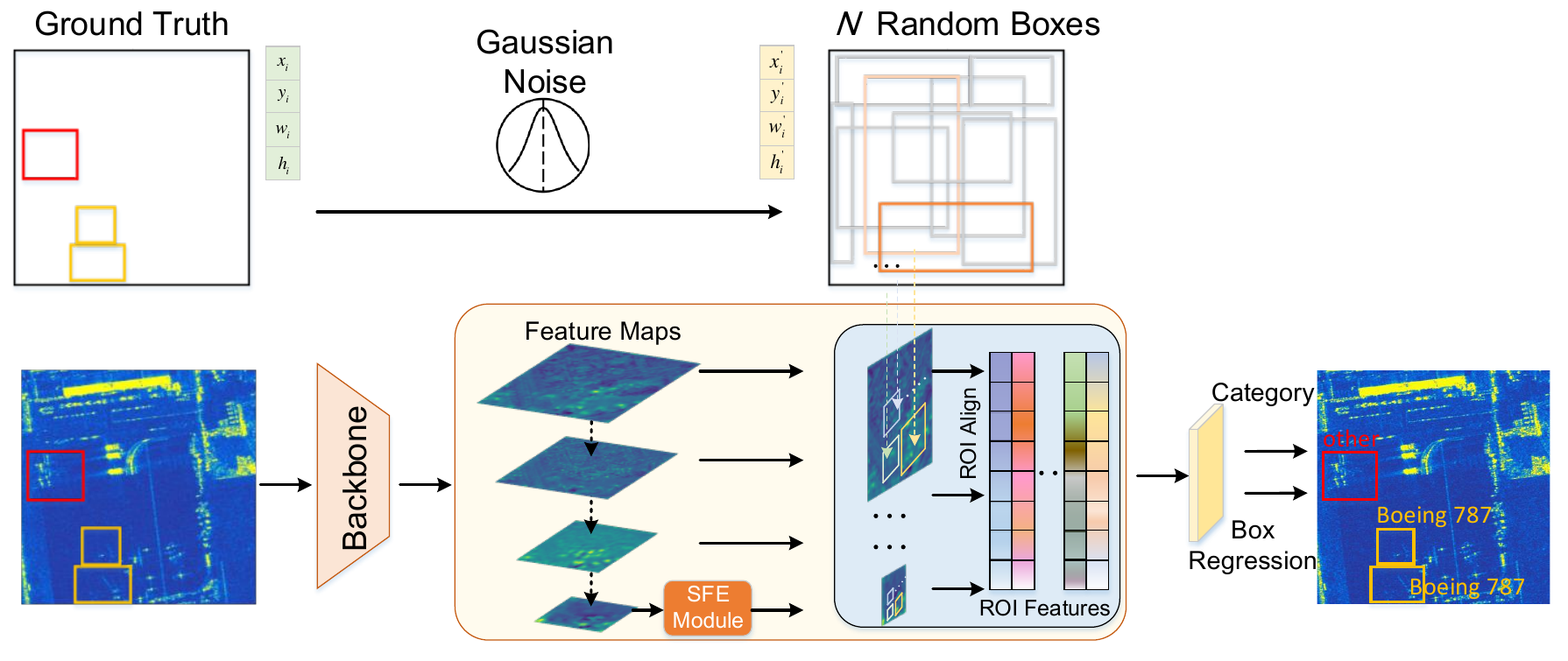}
% \caption{Illustration of the overall framework of our proposed DiffDet4SAR }
   \subfloat{}
  \includegraphics[width=0.35\textwidth]{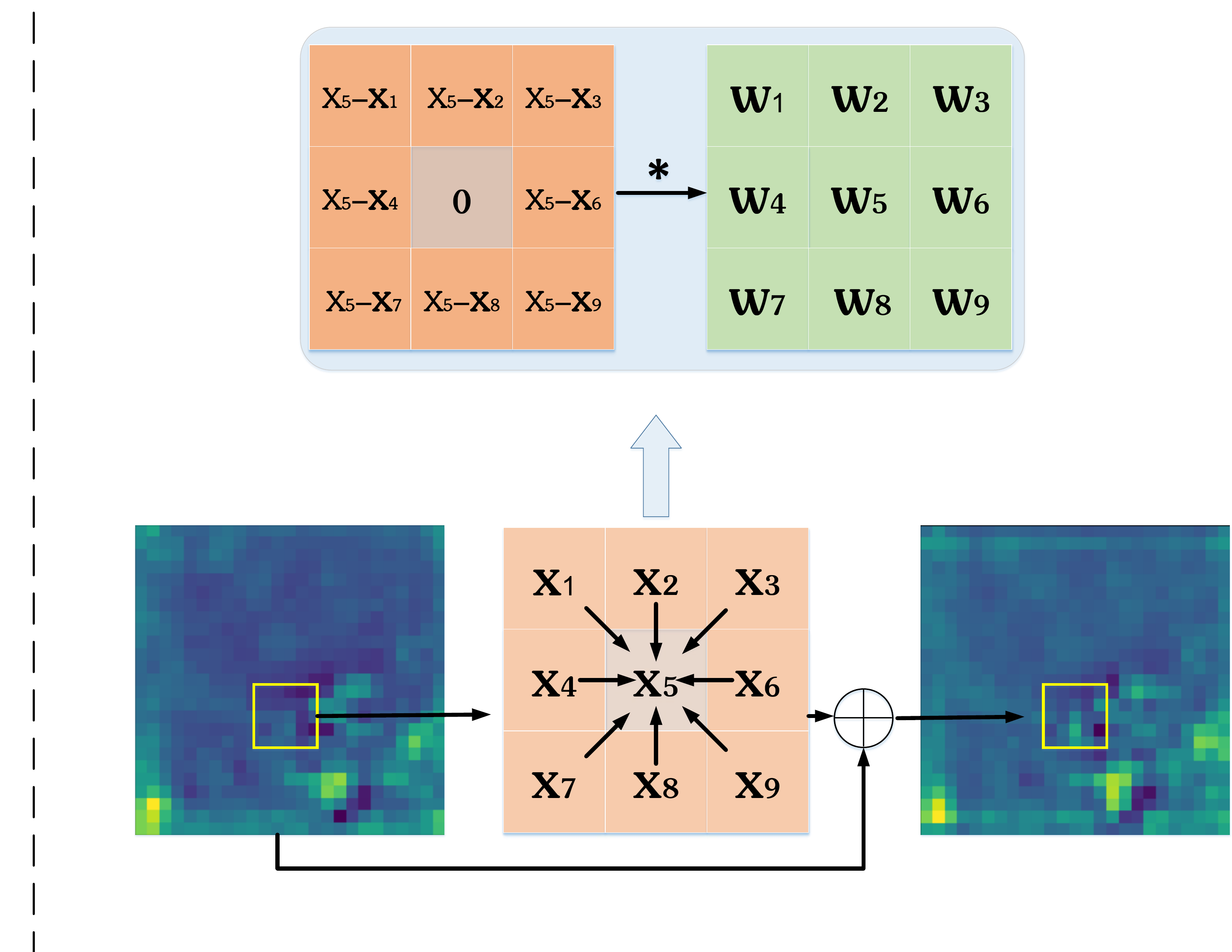}
  % \caption{SFE Module }
   \begin{center}
    	\footnotesize  ~~~~~~~~~(a) DiffDet4SAR\qquad\qquad\qquad\quad~~~~\qquad\qquad\qquad\quad~~~~\qquad\qquad\qquad\quad~~~~\qquad\qquad\qquad  ~~~~\quad~~~~~(b) SFE Module
    \end{center} 

 \caption{ Illustration of the overall framework of our proposed DiffDet4SAR (a) and detailed implementation of Scatterring Feature Enhancement (SFE) module (b). The backbone extracts feature maps from the input SAR image and the SFE module is applied to the high-level semantic feature map. Taking noisy bounding boxes and multi-scale features as input, the detector then predicts the target category, positions (center coordinates), and sizes (widths and heights) of bounding boxes. Furthermore, we design the SFE module to suppress background clutter and enhance the saliency of targets.}
  \label{fig: framework} 
\end{figure*}

\section{Methods}
The overall architecture of the proposed DiffDet4SAR is illustrated in Fig. \ref{fig: framework}. The backbone extracts feature maps from the input SAR image and we employ SFE module on the high-level semantic feature map. Then, taking noisy bounding boxes and multi-scale features as input, the detector predicts the target category, positions (center coordinates), and sizes (widths and heights) of bounding boxes. Details of the diffusion-based aircraft target detection and SFE module are described in Sections II-A and II-B, respectively.

\subsection{Diffusion-based Aircraft Target Detection}
Diffusion models were first introduced in \cite{sohl2015deep} as a class of likelihood-based models inspired by nonequilibrium thermodynamics. Unlike traditional methods that directly model the data distribution, diffusion models are parameterized by a Markov chain that gradually adds noise to the data until the signal is totally destroyed. During the inference stage, a sample from the training distribution can be generated from a randomly sampled Gaussian noise by iteratively applying a reverse diffusion step. Furthermore, diffusion models are trained by optimizing the variational lower bound of the negative log-likelihood of the data. 

Starting with SAR image $x_{0}$, the forward process $q$ adds Gaussian noise $\epsilon$ over $T$ steps according to the noise schedule $\alpha_{t}$, where $t\in [1,2,...,T]$. The forward process is defined as:

\begin{equation}
\begin{aligned}
x_t=\sqrt{\alpha_t} x_{t-1}+\sqrt{1-\alpha_t} \epsilon ; \epsilon \sim \mathcal{N}(\epsilon ; \mathbf{0}, \mathbf{I}),
\end{aligned}
\end{equation}

\begin{equation}
\begin{aligned}
q\left(x_t \mid x_{t-1}\right)=\mathcal{N}\left(x_t ; \sqrt{\alpha_t} x_{t-1},\left(1-\alpha_t\right) \mathbf{I}\right),
\end{aligned}
\end{equation}

\begin{equation}
\begin{aligned}
q\left(x_{1: T} \mid x_0\right)=\prod_{t=1}^T q\left(x_t \mid x_{t-1}\right),
\end{aligned}
\end{equation}
where the noise parameters follow a Gaussian normal distribution with $x_{T}\sim \mathcal{N}(0,I)$.

The reverse process $p$ can be described as:
\begin{equation}
\begin{aligned}
p\left(x_{0: T}\right)=p\left(x_T\right) \prod_{t=1}^T p_\theta\left(x_{t-1} \mid x_t\right).
\end{aligned}
\end{equation}

Each step of the denoising process is learned by a neural network parameterized by $\theta$, and can be simplified as:

\begin{equation}
\begin{aligned}
p_\theta\left(x_{t-1} \mid x_t\right)=\mathcal{N}\left(x_{t-1} ; \mu_\theta\left(x_t, t\right), \Sigma_\theta\left(x_t, t\right)\right).
\end{aligned}
\end{equation}

In this letter, we describe aircraft target detection as the task of generating the position (center coordinates) and size (width and height) of the bounding boxes in the image space. The proposed method does not require prior candidate boxes, alleviating the inaccurate positioning and missed detections caused by the diversity of aircraft sizes and the deviations caused by incorrect manual settings. The learning objective for aircraft target detection is the input-target pairs $(x, b, c)$, where $x$ is the input image, and $b$ and $c$ are the bounding box and class label in SAR image $x$, respectively. Specifically, we represent the $i$-th box in the set as $b^{i}$ = $(c_x^{i},c_y^{i},w^{i},h^{i})$, where $(c_x^{i},c_y^{i})$ are the center coordinates of the bounding box and $(w^{i},h^{i})$ are its width and height. The forward noise process can be defined as:

\begin{equation}
\begin{aligned}
q\left(z_t \mid z_{0}\right)=\mathcal{N}\left(z_t ; \sqrt{\bar{\alpha_t}} z_{0},\left(1-\bar{\alpha_t}\right) \mathbf{I}\right),
\end{aligned}
\end{equation}
which transforms data sample $z_{0}$ into a latent noisy sample $z_{t}$ for $t \in [0,1,...,T]$ by adding noise to $z_{0}$, with $\bar{\alpha_{t}}= \alpha_{1}\cdot \alpha_{2}...\cdot \alpha_{t}$. 
% Considering the diverse scales and large differences of aircraft targets, we introduce diffusiondet with the need for heuristic object priors to perform aircraft target detection.

In our setting, the data sample is a set of bounding boxes $z_0 = b$, where $b\in R^{N\times 4}$ represents a collection of $N$ bounding boxes (the four dimensions represent the center coordinates, width, and height of the boxes). The neural network $f_{\theta } (z_t,t,x)$ is trained to predict $z_0$ and generate the corresponding class label $c$ from the noisy box $z_t$ by minimizing the training objective in terms of $L_{2}$ loss:

\begin{equation}
\begin{aligned}
\mathcal{L}_{\text {train }}=\frac{1}{2}\left\|f_\theta\left(\boldsymbol{z}_t, t\right)-\boldsymbol{z}_0\right\|^2.
\end{aligned}
\end{equation}

As shown in Fig. \ref{fig: framework} (a), during the training phase, Gaussian noise is added to the ground truth box to obtain a noisy box with variance scheduling control. The features from RoI are clipped from the shallow feature map and the enhanced high-level semantic features by these noisy boxes. Finally, these RoI features are sent to the detector, which is trained to predict the ground truth box without noise. During the inference phase, DiffDet4SAR generates bounding boxes by reversing the learned diffusion process, which adjusts the noise prior distribution to the distribution of the learned bounding boxes.

\subsection{Scattering Feature Enhancement Module}
SAR images suffer from severe background interference, which causes aircraft targets to be submerged. To effectively suppress the background clutter, we design the SFE module and innovatively utilize target neighborhood information. The structure of the SFE module is illustrated in Fig. \ref{fig: framework} (b). Specifically, considering the anisotropy of the target and the self-similarity of the background, we introduce the central pixel difference convolution\cite{su2021pixel} to reduce the background scattering intensity and suppress clutter.

% The aircraft dataset presents a challenging scenario with significant background interference, resulting in less prominent targets. To tackle this issue, we propose leveraging context information and neighborhood representation to enhance target visibility and capture detailed features. To achieve this, we introduce the central difference convolution technique\cite{su2021pixel}, which effectively enhances the saliency of targets, suppresses background interference, and improves target localization accuracy. 

The process of center pixel difference convolution (PDC) is similar to that of vanilla convolution, where the original pixels in the local feature map patch are covered by the convolution kernels and replaced by pixel differences in the convolutional operation. The formulations of vanilla convolution and PDC can be expressed as:

% \begin{equation}
% % \begin{aligned}
% y=f(\boldsymbol{x}, \boldsymbol{\theta})=\sum_{i=1}^{k \times k} w_i \cdot x_i, \quad \text { (vanilla convolution) }
% % \end{aligned}
% \end{equation}

% \begin{equation}
% % \begin{aligned}
% y=f(\nabla \boldsymbol{x}, \boldsymbol{\theta})=\sum_{\left(x_i, x_i^{\prime}\right) \in \mathcal{P}} w_i \cdot\left(x_i-x_i^{\prime}\right), \quad \text { (PDC) }
% % \end{aligned}
% \end{equation}

\begin{equation}
\begin{aligned}
& y=f(\boldsymbol{x}, \boldsymbol{\theta})=\sum_{i=1}^{k \times k} w_i \cdot x_i, \quad \text { (vanilla convolution) } \quad \\
& y=f(\nabla \boldsymbol{x}, \boldsymbol{\theta})=\sum_{\left(x_i, x_i^{\prime}\right) \in \mathcal{P}} w_i \cdot\left(x_i-x_i^{\prime}\right), \quad \text { (PDC) } \quad 
\end{aligned}
\end{equation}
where $x_i$ and $x_i^{\prime}$ are the input pixels and $w_i$ is the weight in the $k\times k$ convolution kernel. $\mathcal{P}={(x_1,x_1^{\prime}),(x_2,x_2^{\prime}),...,((x_m,x_m^{\prime}))}$ is the set of pixel pairs picked from the current local patch, and $m \le k \times  k$. Given the self-similarity of the background, using central pixel difference convolution to subtract the weighted surroundings around the target suppresses the background interference while enhancing the saliency of the anisotropic target.

Considering the discrete characteristics of aircraft scattering points, applying central difference convolution to shallow feature maps may cause the model to focus too much on texture and independent scattering points, leading to missed detections and false alarms. To address this issue, we apply central pixel difference convolution to the high-level semantic feature layer. Furthermore, to enhance target saliency, we design a feature fusion module that combines the original features with the clutter-suppressed feature maps to improve the target scattering intensity. This module highlights the target, reduces background interference, and improves detection accuracy.

% 我们对图像经过BackboneHour的特征图的最后一层做中心差分卷积，提高目标显著性，抑制杂波，然后将特征图输入到检测头进行预测。

\section{Experiments}
\subsection{Dataset and Experimental Setup}
\subsubsection{Dataset}
To evaluate the performance of the proposed DiffDet4SAR, we perform extensive experiments using the SAR-AIRcraft-1.0 dataset\cite{zhirui2023sar}, a recently released multi-category SAR aircraft target detection dataset with challenging scenarios. The dataset includes four different scales: 800$\times$800, 1000$\times$1000, 1200$\times$1200, and 1500$\times$1500.  The aircraft targets in the dataset have a wide-ranging size distribution. Some target sizes are less than 50$\times$50, while others exceed 100$\times$100. The specific categories of aircraft targets include A220, A320/321, A330, ARJ21, Boeing737, Boeing787, and other. And \textit{other} represents instances that do not belong to the defined 6 categories. 

\subsubsection{Experimental Setup}
Experiments using the proposed and comparative methods were implemented in PyTorch and executed on a personal computer with an NVIDIA RTX 4090 and 24GB memory. The ResNet50, which had been pre-trained on ImgaeNet data, was used as the backbone of our proposed network. During training, the batch size was set to 8, the initial learning rate was $2.5\times10^{-5}$, and the weight decay was $10^{-4}$. 

\subsubsection{Evaluation Metrics}
% 具体设置的实验参数评价指标等
The average precision (AP), mean average precision (mAP) \cite{niu2023aircraft}, and F1-score at the intersection over union (IoU) thresholds of 0.5 and 0.75 were used to evaluate the performance of different methods.

\begin{table*}[t]
 \centering
 \caption{Comparisons of detection performance on SAR-AIRcraft-1.0\cite{zhirui2023sar}. The results of each category are computed by mAP(\%) and F1-score (\%) with an intersection over the union threshold of 0.5 \textbf{(IoU=0.5)}. The \textbf{best} and \underline{second best} results are shown in \textbf{bold} and \underline{underline}.}
 \label{tab:performance when iou=0.5}
 
\resizebox{0.9\linewidth}{!}{

 \begin{tabular}{cccccccccc}
  \toprule
Method & A330 & A320/321 & A220 & ARJ21 & Boeing737 & Boeing787 & other  &mAP$_{50}$ & F1-score\\  \midrule

Faster RCNN\cite{ren2017fasterrcnn} & 85.0 & 97.2 & 78.5 & 74.0 & 55.1 & 72.9 & 70.1 & 76.1 &73.9\\
Cascade RCNN\cite{cai2018cascade} & 87.4 & 97.5 & 74.0 & 78.0 & 54.5 & 68.3 & 69.1 & 75.7&71.6 \\
Reppoints\cite{yang2019reppoints} & 89.8 & 97.9 & 71.4 & 73.0 & 55.7 & 51.8 & 68.4 &72.6&- \\
% YOLOv6n\cite{li2023yolov6} & 96.2&97.5&\textbf{92.6}&81.4&\textbf{78.3}&80.6&78.9&86.5&81.3\\
SKG-Net\cite{fu2021scattering} & 79.2 &78.4 & 66.7 &64.8 & 65.8 & 69.3 & 71.7 &70.5&63.2 \\
SA-Net\cite{zhirui2023sar} & 88.6 & 94.3 & 80.3 & 78.6 & 59.7 & 70.8 & 71.3 & 77.7&- \\
MLSDNet\cite{chang2023mlsdnet} & 91.5 & 96.9 & \textbf{85.1} & 83.2 & \underline{71.7} & 72.1 & 78.4 & 82.7&77.9\\
DiffusionDet\cite{chen2023diffusiondet} & \underline{95.4} & \underline{98.1} & 80.8 & \underline{84.2} &70.9 & \underline{91.4} & \textbf{86.4} & \underline{86.6}&\underline{81.4} \\
\rowcolor{LightCyan}DiffDet4SAR(Ours) & \textbf{97.1} & \textbf{99.4} & \underline{82.3}& \textbf{87.2} &\textbf{72.8} & \textbf{93.3} & \underline{85.9} & \textbf{88.4}&\textbf{83.7} \\
  \bottomrule
 \end{tabular}}
\end{table*}

\begin{figure}[t]
  \centering
  \subfloat{}
  {\includegraphics[width=0.5\textwidth]{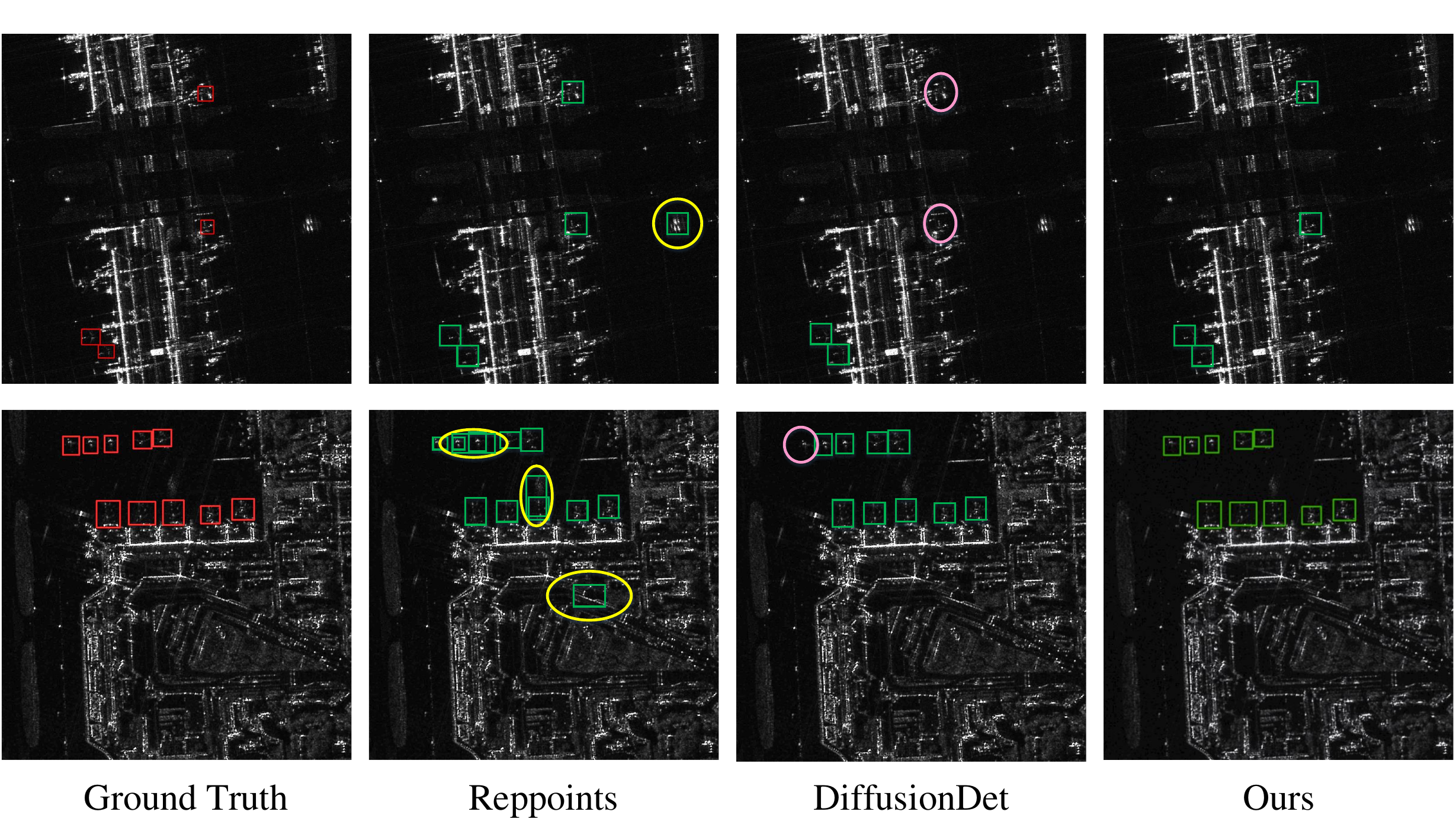}}
  \caption{Visualization of detection results under different methods on randomly selected images. Red boxes represent ground truth, and green boxes indicate detection results. Yellow ellipses indicate false positives, and the purple ellipses represent instances of false negatives.}

  \label{fig:visualization_of_detection_results}
\end{figure}

\subsection{Comparison with the State-of-the-Art Methods}
 We compared our method with six state-of-the-art detection methods, namely Faster RCNN\cite{ren2017fasterrcnn}, Cascade RCNN\cite{cai2018cascade}, Reppoints\cite{yang2019reppoints}, SKG-Net\cite{fu2021scattering}, SA-net\cite{zhirui2023sar}, and MLSDNet\cite{chang2023mlsdnet}. These detection methods include both single-stage and two-stage methods, as well as anchor-based and anchor-free methods.

% YOLOv6n\cite{li2023yolov6},

% 定量分析
\subsubsection{Quantitative Results}
% 以自己为中心

Table \ref{tab:performance when iou=0.5} presents the quantitative detection results from different methods.  It can be observed that our DiffDet4SAR achieves the optimal mAP$_{50}$ value and F1-scrore, outperforming the state-of-the-art SAR detection method (MLSDNet) by 5.7\% in mAP$_{50}$ and 5.8\% in F1-score. This is attributed to our utilization of random noise bounding boxes. Compared with traditional fixed bounding boxes, the random noisy bounding boxes based on diffusion models are not limited by size or aspect ratio, allowing better adaptation to the size variations of aircraft. This results in improved adaptability and robustness of the proposed DiffDet4SAR. Furthermore, introducing the SFE module enables our DiffDet4SAR to outperform the DiffusionDet by 1.8\% in mAP$_{50}$ and 1.7\% in F1-score. The performance improvement is particularly noticeable for small-size targets, with an enhancement of 3\% in mAP$_{50}$ for ARJ21 aircraft. These
results verify that our proposed SFE module
effectively overcomes background interference and significantly enhances the saliency of targets, particularly for difficult categories.

% 一般用50，但是75也可以作为更严格的对比
We also used the more stringent mAP$_{75}$ metric to evaluate the competitiveness of the proposed DiffDet4SAR, as shown in Table \ref{tab:performance when iou=0.75}. With the fusion of global contextual features and scattering information, our DiffDet4SAR achieves a mAP$_{75}$ score of 68.2\%, surpassing that of the best SAR target detection method (SA-Net) by approximately 6\%. Additionally, the SFE module based on the characteristics of aircraft targets in SAR images plays a significant role, enabling our proposed DiffDet4SAR to outperform the DiffusionDet method by 2\% in terms of mAP$_{75}$. Moreover, there are certain differences in detection accuracy for each category. For example, compared withother categories, the A320/321 category gives the highest mAP$_{50}$ and mAP$_{75}$ detection results across different methods. This is mainly because A320/321 aircraft have special dimensions, with a body length of over 40m, making it easily distinguishable. For certain types of targets, such as ARJ21 and A220 aircraft, the detection accuracy is lower due to their smaller size, which makes it difficult to sufficiently capture detailed features.

\begin{table}[t]
 \centering
 \caption{Comparisons of detection performance on SAR-AIRcraft-1.0\cite{zhirui2023sar}. The results of each category are computed by mAP(\%) with an intersection over the union threshold of 0.75 \textbf{(IoU=0.75)}. The \textbf{best} and \underline{second best} results are shown in \textbf{bold} and \underline{underline}.}
 \label{tab:performance when iou=0.75}
 \small
\setlength{\tabcolsep}{1mm}{
 % \begin{tabular}{@{}ccccccc@{}}
  \begin{tabular}{ccccccc}
  \toprule

Category &\thead{Faster\\RCNN}& \thead{Rep-\\points} & SKG-Net & SA-Net & \thead{Diffusion-\\Det} &\thead{DiffDet4SAR\\(Ours)} \\ 
\midrule

A330 & 85.0  & 66.4 & 66.4 & \underline{88.6} &88.3 & \textbf{89.6}\\
A320/321 & \underline{87.7} & 84.9 & 49.6 & 86.6&83.0& \textbf{88.4} \\
A220 & \textbf{58.7}& 49.4 & 29.8 &50.6& \underline{55.0} &  50.6\\
ARJ21 & 55.2  & 50.9 &37.7 &\underline{56.0}& \textbf{59.7} &\underline{56.0} \\
Boeing737 & 42.8 & 36.6 & 48.7 & 41.8&\underline{60.1}&\textbf{64.7} \\
Boeing787 & 60.5 & 41.8 & 51.6 &60.4 &\underline{77.0} &\textbf{77.1}\\
other & 45.4  & 43.1 & 41.1 & 47.7 & \underline{48.6}&\textbf{49.7}\\
 \rowcolor{LightCyan}mAP$_{75}$ & 62.2 & 53.3 & 46.4 &62.8 &\underline{66.1} &\textbf{68.2} \\
  % \cellcolor{LightCyan} mAP$_{70}$ & \cellcolor{LightCyan} 62.2 & 58.9 & 53.3 & 46.4 &62.8 & \textbf{68.2} \\
  \bottomrule
 \end{tabular}}
\end{table}

\begin{table}[t]
 \centering
 \caption{Ablation experiments to determine the effectiveness of SFE (/\%). Using PDC on p5 and integrating vanilla feature with PDC$_{p5}$  works \textbf{best}.}
 \label{tab:vallina_CPDC}
 \small
\setlength{\tabcolsep}{0.6mm}{
 \begin{tabular}{cccccccccc}
  \toprule
Vanilla& PDC$_{p3}$& PDC$_{p4}$&PDC$_{p5}$& mAP & mAP$_{50}$ & mAP$_{75}$&mAP$_{s}$&mAP$_{m}$&mAP$_{l}$\\\midrule
\CheckmarkBold & & & &59.7 & 86.6&66.9&13.6&57.3&55.3\\
 &\CheckmarkBold & & &59.8 & 87.6& 67.2&30.0 &57.8 &55.2 \\
 & &\CheckmarkBold & &59.9 &87.2 &66.8 &20.0 &57.5 &55.5 \\
&&&\CheckmarkBold &60.1 &87.4 &66.0 & 10.0&58.1&55.6\\
 & \CheckmarkBold&\CheckmarkBold& \CheckmarkBold&60.2 &87.6  &67.0  & 25.2 &58.1 &56.2 \\ 
 \Xhline{0.5pt}
\CheckmarkBold&\CheckmarkBold && &60.2 &88.1 &68.1 & 20.0&58.1&56.2\\
\CheckmarkBold& &\CheckmarkBold& &60.3 &87.7 & 66.5 & 30.0 &58.1 &56.4 \\
\rowcolor{LightCyan} \CheckmarkBold& & &\CheckmarkBold &60.6 & 88.4&68.2&30.0&59.0&56.6\\

\CheckmarkBold& \CheckmarkBold&\CheckmarkBold& \CheckmarkBold& 60.5& 88.0 & 68.2 & 30.0 & 58.1&56.5 \\
  \bottomrule
 \end{tabular}}
\end{table}

\begin{figure}
  \centering           
\includegraphics[width=0.5\textwidth]{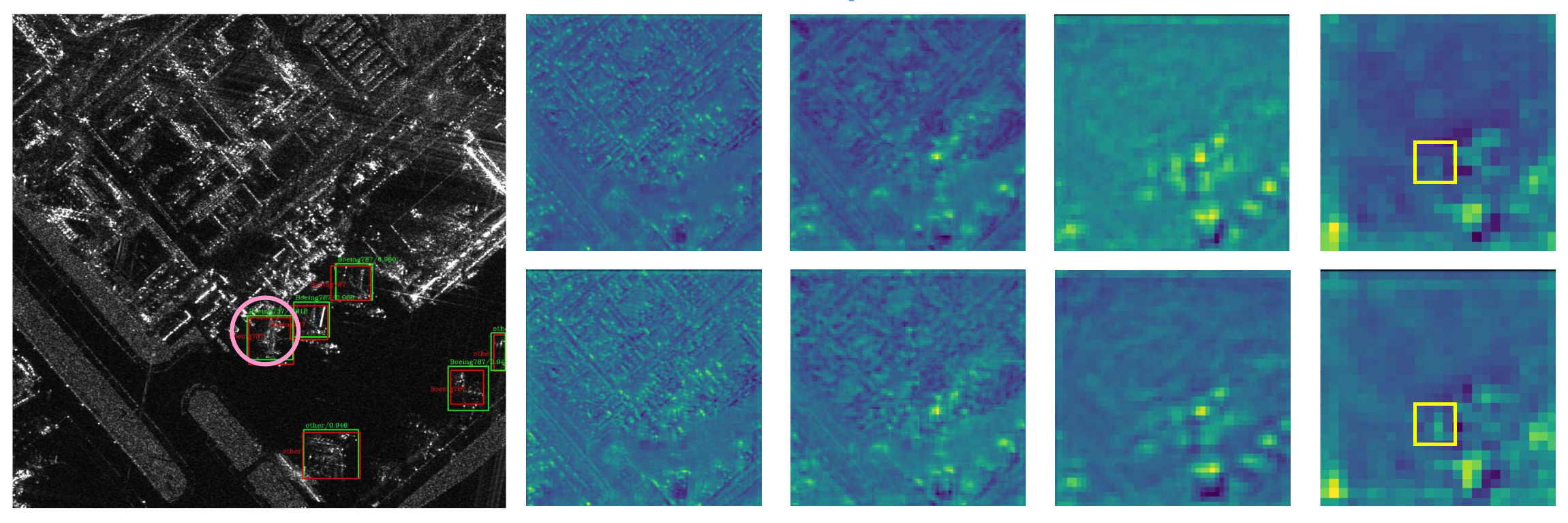}
  \caption{The influence of proposed SFE module on feature maps of different depths. Results show the feature maps without (top row) and with the SFE module (bottom row). The feature maps become deeper from left to right, i.e., from p2 to p5. Red boxes represent the ground truth, and green boxes indicate the detection results. The purple ellipse represents instances of false negatives without using the SFE module in feature layer p5.}
  \label{fig: SFE_comparision} 
\end{figure}

\subsubsection{Qualitative Results}
Fig. \ref{fig:visualization_of_detection_results} illustrates the detection results given by different methods. When there are few instances in the SAR image, DiffDet4SAR can accurately locate and detect the targets, with the predicted boxes closely aligned with the ground truth boxes. For SAR images with densely packed instances, DiffDet4SAR is still able to accurately detect the positions and number of airplane targets, with minimal deviation between the predicted and ground truth boxes. That is the result of random sampling from Gaussian noise, which obtains a diverse range of sizes and aspect ratios and effectively increases the target coverage. 
Furthermore, the proposed DiffDet4SAR outperforms DiffusionDet in the case of dense targets. This is because SFE module highlights target saliency while suppressing the visual similarity between targets and surrounding clutter, thus reducing missed detections.

\subsection{Ablation Study}

% 通过融合最后一层特征和经过中心差分卷积之后的特征，可以增强模型对于局部细节和上下文信息的捕获能力，从而提升模型的检测效果。这种融合方式可能有助于模型更好地理解输入数据的结构和语义信息，进而提高其性能表现。
\subsubsection{Effectiveness of SFE}

To investigate the effectiveness of our proposed SFE module, we compare different feature extraction strategies in Table \ref{tab:vallina_CPDC}. When evaluating DiffDet4SAR without PDC or feature fusion, there are slight benefits in terms of mAP. Moreover, our DiffDet4SAR attains remarkable gains when equipped with both vanilla and PDC$_{p5}$ features. These experiments together verify the necessity of the SFE module. Fig. \ref{fig: SFE_comparision} illustrates that using central PDC in deep semantic feature maps effectively enhances the intensity of targets. By integrating the features from feature layer p5 with those obtained through central PDC, the model realized an enhanced ability to capture local details and contextual information in SAR images, thereby improving its detection performance. The SFE module helps the model to better understand the scattering structures and semantic information present in SAR images, thus improving overall performance.

% 模型在N的数量更多的时候表现更好，这可能是因为当训练时N的数量越多，模型能够学习到目标的更多特征信息，有利于检测结果的提升。
\subsubsection{Signal Scaling}
The signal scaling factor controls the signal-to-noise ratio (SNR) of the diffusion process. Results in Table \ref{tab:signal_scale}
demonstrate that a scaling factor of 1.0 achieves the optimal mAP value, outperforming the standard value of 2.0 used
in DiffusionDet\cite{chen2023diffusiondet}. This is because SAR images contain significant clutter interference and a large scaling factor submerges the target in heavy background clutter. However, if the SNR is too small, such as 0.1, the feature maps contain insufficient target information, degrading the detection performance.

\begin{table}[t]
 \centering
 \caption{Signal scale. Normal scaling factor improve detection performance(/\%).}
 \label{tab:signal_scale}
 \small

\resizebox{0.95\linewidth}{!}{
 \begin{tabular}{ccccccc}
  \toprule
scale& mAP & mAP$_{50}$ & mAP$_{75}$ &mAP$_{s}$&mAP$_{m}$&mAP$_{l}$\\\midrule
0.1 &49.6 &73.1 &56.3& 30.0&45.0&53.4\\
 \rowcolor{LightCyan} 1.0&60.6 & 88.4&68.2&30.0&58.1&56.6\\
2.0 &60.0 & 86.8&67.3&13.6&57.8&55.3\\
3.0 &60.1 &87.4 &66.0 & 10.0&58.1&55.3\\
  \bottomrule
 \end{tabular}}
\end{table}

\subsubsection{Matching between N$_{train}$ and N$_{eval}$}

The proposed network has the appealing property of accepting an arbitrary number of random boxes. To investigate the impact of the number of training boxes on the inference performance, we trained our model with $N_{train}\in [{100, 300, 500}]$ random boxes, and then evaluated each of these models with $N_{eval}\in[{100, 300, 500,1000}]$, as summarized in Table \ref{tab:box_numbes}. Regardless of the value of $N_{train}$, the accuracy remains steady. $N_{train}$ has a greater impact on the detection results than $N_{eval}$ and the model tends to perform better with high values of $N_{train}$. This is because, when there are more random boxes during training, the model can learn more texture and semantic feature information about the targets, which is beneficial for improving the detection results. During the validation stage, there are typically fewer targets present in SAR images than the value of $N_{eval}$. As a result, many boxes that do not contain targets are suppressed and discarded, leading to a relatively small improvement in detection results as $N_{eval}$ increases.

\begin{table}[t]
 \centering
 \caption{Matching between training and inference box numbers on SAR-AIRcraft-1.0\cite{zhirui2023sar} (mAP/\%). DiffDet4SAR decouples the number of boxes during the training and inference stages and works well with flexible combinations.}
 \label{tab:box_numbes}
 \small
\setlength{\tabcolsep}{2.5mm}{
 \begin{tabular}{c|cccc}
 \Xhline{0.8pt}
\diagbox[width=5em,trim=l]{train}{eval}& 100 & 300 & 500 & 1000\\  \Xhline{0.8pt}
100 &59.9 &59.7&59.7 &59.7\\
300&59.6 & 60.0&59.6&59.6\\
 \rowcolor{LightCyan} 500 & 60.6& 60.6&60.6&60.6\\

 \Xhline{0.8pt}
 \end{tabular}}
\end{table}

\section{Conclusion}
In this letter, we propose a diffusion-based aircraft target detection
network for SAR images named DiffDet4SAR, presentING a first study that introduces diffusion models to SAR target detection. We formulated aircraft target detection as a spatial generation task of the position and size of bounding boxes. Furthermore, to address the discrete nature of scattered points from aircraft targets and strong background interference, we designed the SFE module to capture texture details and semantic information as a means of enhancing target saliency and suppressing background clutter. The experimental results presented herein have demonstrated the effectiveness of the proposed DiffDet4SAR method. 

% We believe this diffusion-based norm will provide valuable insights for SAR image detection tasks.

%  To the best of our knowledge, this is the first attempt to apply diffusion models to the problem of SAR target detection, providing a promising new paradigm for SAR image interpretation.

% We design the scattering feature enhancement (SFE) module, which addresses the problems of discrete scattering points of aircraft targets and severe background interference. The SFE module effectively reduces the scattering intensity of background clutter and highlights targets.

\bibliographystyle{ieeetr}
\bibliography{./ref}

\end{document}